\documentclass[useAMS]{mn2e}

\usepackage{epsfig}

\newcommand{\be}{\begin{eqnarray}}
\newcommand{\ee}{\end{eqnarray}}
\def\lsim{\,\lower2truept\hbox{${< \atop\hbox{\raise4truept\hbox{$\sim$}}}$}\,}
\def\gsim{\,\lower2truept\hbox{${> \atop\hbox{\raise4truept\hbox{$\sim$}}}$}\,}

\title[Metals, dust and the CMB]{Metals, dust and the cosmic microwave background: fragmentation of high-redshift star-forming clouds}
\author[Raffaella Schneider and Kazuyuki Omukai]{Raffaella Schneider$^{1}$\thanks{E-mail:
raffa@arcetri.astro.it} and Kazuyuki Omukai$^{2}$\\
$^{1}$ INAF/Osservatorio Astrofisico di Arcetri, Largo Enrico Fermi 5, 50125 Firenze, Italy\\
$^{2}$ National Astronomical Observatory of Japan,  Mitaka, Tokyo 181-8588, Japan}

\begin{document}

\date{September 2009}

\pagerange{\pageref{firstpage}--\pageref{lastpage}} \pubyear{2009}

\maketitle

\label{firstpage}

\begin{abstract}

We investigate the effects of the Cosmic Microwave Background (CMB) radiation field on
the collapse of prestellar clouds. Using a semi-analytic model to follow the 
thermal evolution of clouds with varying initial metallicities and dust contents 
at different redshifts, we study self-consistently the response of the mean Jeans mass 
at cloud fragmentation to metal line-cooling, dust-cooling and the CMB.

In the absence of dust grains, at redshifts $z \le 10$ moderate characteristic masses 
(of 10s of $M_{\odot}$) are formed when the metallicity is $10^{-4} Z_{\odot} \le Z \le 
10^{-2.5} Z_{\odot}$; at higher metallicities, the CMB inhibits fragmentation and only 
very large masses (of $\sim$ 100s of $M_{\odot}$) are formed. These effects become even 
more dramatic at $z > 10$ and the fragmentation mass scales are always $\ge$ 100s of $M_{\odot}$, 
independent of the initial metallicity. 

When dust grains are present, sub-solar mass fragments are formed at any redshift 
for metallicities $Z \ge 10^{-6} Z_{\odot}$ because dust-cooling remains relatively 
insensitive to the presence of the CMB. When $Z > 10^{-3} Z_{\odot}$, heating of dust 
grains by the CMB at $z \ge 5$ favors the formation of larger masses, which become 
super-solar when $Z \ge 10^{-2} Z_{\odot}$ and $z \ge 10$. 
Finally, we discuss the implications of our result for the interpretation 
of the observed abundance patterns of very metal-poor stars in the galactic halo.  

\end{abstract}


\section{Introduction}

The origin of the stellar Initial Mass Function (IMF) and its
variation with cosmic time or with diverse environmental conditions 
still lack a complete physical interpretation. Observationally, 
the present-day stellar IMF appears to have an almost universal profile, 
characterized by a power-law at large masses and flattening 
below a characteristic mass of $M_{\rm ch} \approx 1 M_\odot$ with a plateau 
which extends down to the brown dwarf regime (see Chabrier 2003 and
references therein). 
This mass scale, which represents the "typical" outcome 
of the star formation process, appears to be remarkably uniform in diverse 
environments 
where the initial conditions for star formation 
can vary considerably (see Elmegreen, Klessen \& Wilson 2008). 

Among the many proposed explanations, the origin of the characteristic 
stellar mass and the broad peak of the IMF are best attributed to 
gravitational fragmentation, which sets the mean 
Jeans mass at cloud fragmentation (see Bonnell, Larson \& Zinnecker 
2007 for a comprehensive review),
\begin{equation}
M_{\rm frag} = \rho \lambda_J^3 \sim 700 M_{\odot} \left(\frac{T_{\rm frag}}{200~\rm{K}}\right)^{3/2} \left(\frac{n_{\rm frag}}{10^4 \rm{cm}^{-3}}\right)^{-1/2} 
\label{eq:mfrag}
\end{equation}
where the Jeans length is defined as,
\[
\lambda_J = \left(\frac{\pi k T}{\mu m_H G \rho}\right)^{1/2}.
\]
\noindent
The process of gravitational fragmentation can be investigated using 
thermal physics within the framework of the classical Jeans criterion: 
fragmentation occurs at the end of an efficient cooling phase, when the 
temperature decreases for increasing density and the effective adiabatic 
index switches from $\gamma < 1$ to $\gamma > 1$ 
(the pressure $p \propto \rho^{\gamma}$, where $\rho$ is the density); 
the typical fragmentation mass-scale is given by the thermal Jeans mass 
at that epoch (Larson 2005; Schneider et al. 2002). The appearance of a 
characteristic mass at the inflection point in the equation of state has 
been also demonstrated numerically in 3D simulations by Jappsen et al. 
(2005), supporting the idea that the distribution of stellar masses
depends mainly on the thermodynamical state of the star-forming gas.  

In the physical conditions which apply to star forming regions in the 
present-day Universe, the application of the above criterion leads to 
a characteristic fragmentation mass of $M_{\rm frag} \sim 2 M_\odot$. 
About a half of this gas is expected to be ejected by a protostellar outflow 
without accreting (e.g. Matzner \& McKee 2000). 

If we apply the same criterion to primordial star forming regions we find a 
fragmentation mass of order $M_{\rm frag} \sim 700 M_\odot$, which corresponds to the Jeans 
mass at the temperature and density where cooling by molecular hydrogen 
becomes less efficient due to the NLTE-LTE transition in the level 
populations of the molecule. These clumps are the immediate progenitors of 
the stars, in agreement with the results found by 3D numerical simulations
(Abel, Bryan \& Norman 2002; Bromm, Coppi \& Larson 2002; Yoshida, Omukai \&
Hernquist 2008). 

Thus, between the primordial and the present-day conditions 
there is a clear transition in fragmentation scales which is believed to be
mainly driven by the progressive enrichment of star forming clouds with 
metals (Bromm et al. 1999; Omukai 2000; Schneider et al. 2002; 
Bromm \& Loeb 2003; Santoro \& Shull 2006; Smith, Sigurdsson \& Abel 2008) and dust grains 
(Schneider et al. 2003; Omukai et al. 2005; Tsuribe \& Omukai 2006; 
Clark, Glover \& Klessen 2008). 

Note, however, that the presence of gas-phase metals can not by itself lead to solar or sub-solar
characteristic masses. In fact, metal line-cooling is efficient in the initial 
evolution of collapsing gas clouds, for densities in the range 
$10^4 - 10^6~$cm$^{-3}$, and induces fragmentation into relatively large clumps, 
with characteristic fragment masses of a few 10s of $M_\odot$ up to initial gas 
metallicities $Z \lsim 10^{-2} Z_{\odot}$ (Schneider et al. 2006).
Conversely, the formation of solar or sub-solar mass fragments might operate already at 
metallicities $Z_{\rm cr} = 10^{-6} Z_{\odot}$ in the presence of dust grains, which
provide an efficient source of cooling at densities $\gsim 10^{13}$cm$^{-3}$.
Since dust can be promptly synthesized in the ejecta of primordial supernovae (Todini \& Ferrara 2001;
Nozawa et al. 2003; Schneider, Ferrara \& Salvaterra 2004; Bianchi \& Schneider 2007), 
it is likely that its contribution to thermal and fragmentation history of collapsing clouds at 
high redshifts had been relevant.

An additional redshift-dependent effect is induced by the Cosmic Microwave Background (CMB) 
which sets an effective temperature floor increasing with redshift. 
Clarke \& Bromm (2003) have proposed a simple formula
for the characteristic stellar mass which is identified by the Jeans mass fixed jointly
by the gas temperature and the pressure set by the weight of the overlying baryons as the
gas collapses in the parent dark matter halo. In this simple model, they do not follow the
gas thermal evolution and the temperature is taken to be the maximum value between the
CMB floor and either $200$~K for H$_2$-cooling or $10$~K for CO-cooling. In the latter case, 
the characteristic stellar mass is set by the CMB temperature at all redshifts
$z > 2.7$ and it is predicted to be $> 10 M_{\odot}$ at $z > 10$. 

The CMB influences the thermal evolution of collapsing gas clouds in two possible 
ways: (i) it affects the atomic and molecular line level populations
(ii) it heats the dust grains. In Omukai et al. (2005), we have 
estimated the impact of the CMB temperature floor at $z = 20$ on the evolution 
of prestellar clouds with varying initial metallicity. It was shown that the 
CMB affects the cloud evolution at low temperatures, which are reached for 
relatively large values of the initial metallicity, $Z > 10^{-3} Z_\odot$.

Smith et al. (2009) have investigated the effects of the CMB on 
the collapse of prestellar clouds with varying initial metallicities by means of numerical
simulations. They confirm that due to metal line-cooling, the gas becomes thermally coupled to 
the CMB for $Z > 10^{-2.5} Z_\odot$ early in the evolution of the clouds when the density 
is still low. Note that a similar effect has been recently found by Jappsen et al. (2009)
using 3D simulations starting from different initial conditions, namely protogalaxies 
forming within a previously ionized HII region that has not yet had time to cool and recombine. 
Even in the latter models, the gas cools rapidly to the CMB floor and no fragmentation occurs
during the collapse for metallicities as high as $0.1 Z_{\odot}$. 

Smith et al. (2009) conclude that the CMB inhibits 
fragmentation at large metallicities and that there were three distinct 
modes of star formation at high redshift: a ‘primordial’ mode, producing very massive stars 
at very low metallicities ($Z < 10^{-3.75} Z_\odot$); a CMB-regulated mode, 
producing moderate mass (10s of $M_\odot$) stars at high metallicites 
($Z > 10^{-2.5} Z_\odot$ at redshift $z \sim 15-20$); and a low-mass (a few $M_{\odot}$) mode 
existing between those two metallicities. As the Universe ages and the CMB temperature decreases, 
the range of the low-mass mode extends to higher metallicities, eventually becoming 
the only mode of star formation. 

In their simulations, however, Smith et al. (2009) consider the thermal 
evolution in the presence of molecular coolants (H$_2$ and HD) and 
gas-phase metals only. They do not include H$_2$ formation heating, 
which affects the thermal evolution at moderate densities 
$\geq 10^8$cm$^{-3}$ in the primordial gas and at lower densities in the metal-enriched gas. 
Moreover, they do not take into consideration the possible presence of dust grains 
which significantly affect the gas thermal evolution even at very low metallicities 
via cooling by thermal emission and efficient H$_2$ formation on their surfaces 
(Schneider et al. 2003; Omukai et al. 2005; Schneider et al. 2006).  

From an observational point of view, the number and properties of 
very metal-poor (VMP) stars in the Galactic halo with [Fe/H] $<-2$ 
seem to require a higher characteristic stellar mass in the past, with a 
redshift-modulation consistent with that expected by the CMB 
(Hernandez \& Ferrara 2001). 
Recent analyses of the surface elemental composition of very 
metal-poor halo stars (Cayrel et al. 2004) show that
the observed abundance pattern appears consistent with the predicted yields of 
Pop III stars with characteristic mass of $\approx 10 M_{\odot}$ (Heger \& Woosley 2008). 
Similar conclusions have been derived analysing the fraction of carbon-enhanced stars 
(CEMP, Tumlinson 2007; Komiya et al 2007). CEMP stars are a subset of metal-poor stars that show 
enhanced carbon-to-iron abundances ([C/Fe] $> 1$) (Beers \& Christlieb 2005). 
These stars are thought to be metal-poor low-mass stars ($< 0.8 M_\odot$) that have acquired 
C enhancements at their convective surfaces by capturing the C-rich ejecta of an AGB companion 
($1.5 M_\odot <  M_\star < 8 M_\odot$). 
Because the binary system that produces a CEMP star requires 
both a low-mass star and an intermediate mass star, the incidence and chemical abundance 
signatures of CEMP stars reflects the underlying IMF in the range $1 M_\odot < M_\star < 8 M_\odot$ 
(Lucatello et al. 2005; Tumlinson 2007; Komiya et al. 2007).

On the basis of these considerations, Tumlinson (2007) suggests that the CMB
is shaping the characteristic mass of the IMF and proposes the following 
parametrization, 
\begin{equation}
M_{\rm ch}/M_\odot = 1.06 M_\odot \left(\frac{{\rm max}[(1+z)2.73~{\rm K}, 8~{\rm K}]}{10~{\rm K}}\right)^{1.7},
\end{equation}
\noindent
which implies that $M_{\rm ch} = 0.72, 6.87$ and $20.63 M_\odot$ at $z = 0, 10$ and $20$.

The question is whether this observationally motivated CMB-regulated IMF can be reconciled with 
theoretical models for the thermal evolution and fragmentation of star-forming clouds at very 
low metallicities. 

Here we intend to revisit the role of the CMB in the evolution of protostellar clouds
investigating in a self-consistent way the response of the fragmentation mass scale
to metal line-cooling, dust cooling, and the CMB.

The paper is organized as follows: in section \ref{sec:model} we present the model 
adopted to follow the thermal and chemical evolution of collapsing protostellar clouds,
in section \ref{sec:results} we illustrate the main results of the analysis, and in
section \ref{sec:conclusions} we discuss their implications. 

\section{The model}
\label{sec:model}

The thermal and chemical evolution of star forming gas clouds is studied
using the model developed by Omukai (2000) and improved in
Omukai et al. (2005). The collapsing gas clouds are described by a one-zone
approach where all physical quantities are evaluated at the center as a
function of the central density of hydrogen nuclei, $n_{\rm H}$. The temperature
evolution is computed by solving the energy equation,

\begin{equation}
\frac{de}{dt}=-p  \,\frac{d}{dt} \,\frac{1}{\rho} - \Lambda_{\rm net}
\label{eq:energy}
\end{equation}
where the pressure, $p$,  and the specific thermal energy, $e$, are
given by,

\be
p = \frac{\rho  \,k  \,T}{\mu  \,m_{\rm H}},
\ee

\be
e=\frac{1}{\gamma_{\rm ad}-1} \frac{k \,T}{\mu  \,m_{\rm H}},
\ee
\noindent
and $\rho$ is the central density, $T$ is the temperature, $\gamma_{\rm ad}$ is the adiabatic exponent,
$\mu$ is the mean molecular weight and $m_{\rm H}$ is the mass of hydrogen nuclei.
The terms on the right-hand side of the energy
equation are the compressional heating rate,
\be
\frac{d\rho}{dt} = \frac{\rho}{t_{\rm ff}}  \qquad \mbox{with} \qquad t_{\rm ff} = \sqrt{\frac{3 \pi}{32 G \rho}},
\ee

\noindent
and the net cooling rate, $\Lambda_{\rm net}$, which consists
of three components,

\be
\Lambda_{\rm net} = \Lambda_{\rm line} + \Lambda_{\rm cont} + \Lambda_{\rm chem}.
\ee
\noindent
The first component, $\Lambda_{\rm line}$, represents the cooling rate due to the emission of line radiation,
which includes molecular line emission of H$_2$, HD, OH, H$_2$O and CO, and atomic fine-structure line emission
of CI, CII and OI. Following Omukai et al. (2005), H$_2$ collisional transition rates and HD parameters are
taken from Galli \& Palla (1998). The second component, $\Lambda_{\rm cont}$, represents the cooling rate due to
the emission of continuum radiation, which includes thermal emission by dust grains and H$_2$ collision-induced
emission. The last term, $\Lambda_{\rm chem}$, indicates the cooling/heating rate due to chemical reactions.
When the gas cloud is optically thick to a specific cooling radiation, the cooling rate is correspondingly
reduced by multiplying by the photon escape probability. Unless otherwise stated, the treatment of these processes
are the same as in Omukai et al. (2005), to which we refer the interested reader for further details. 

In particular, we adopt the reduced chemical model for low-metallicity clouds described in Section 3.2 of Omukai et al. (2005). This
model reproduces the temperature evolution and the abundances of the dominant chemical species following a network
of 54 reactions (instead of the full 500 reactions of the original model) for 26 species (instead of 50).   

In the present analysis, we are interested in investigating the effects of the CMB in the thermal evolution of
the clouds at different metallicities. The CMB enters in our model in two ways: first through the
level population of atomic/molecular lines, and second through the heating of dust grains. 
These effects prevent the gas and dust temperature from falling below the radiation temperature. 
In addition, the higher dust temperature reduces the H$_2$ formation rate by surface reactions. 
The former effect is included in calculating the level populations of atomic and molecular species as in 
Omukai (2001, see Appendix B). The latter effects are included through the energy-balance equation of grains, 

\begin{equation}
4 \, \sigma  \, T_{\rm gr}^4  \,\kappa_{\rm gr}  = \Lambda_{\rm gas \rightarrow dust} + 4 \, \sigma  \, T_{\rm rad}^4  \,\kappa_{\rm gr}
\end{equation}

\noindent
where the left-hand side is the cooling rate of grains and the right-hand side
is the heating rate by collisions with gas particles and by the CMB (see Omukai et al. 2005), 
and we have made the approximation that the Planck mean opacity for absorption is the same as for emission.

For gas at solar metallicity, the abundances of carbon and oxygen are $y_{\rm C}=0.927 \times 10^{-4}$
and $y_{\rm O} = 3.568 \times 10^{-4}$. The mass fraction of dust grains is $0.939 \times 10^{-2}$
below the water-ice evaporation temperature, and it decreases for higher temperatures as each grain
component evaporates (Pollack et al. 1994). For lower metallicity gas, these values are reduced 
proportionally and we denote the corresponding abundances relative to the solar value as, 
[M/H]$ = {\rm log}_{10}(Z/Z_{\odot})$. 
Strictly speaking, we do not expect the regions enriched by the first supernovae to have a solar
abundance pattern and the same dust grains present in the local neighborhood.
Yet, a more detailed analysis which self-consistently accounts for the elemental abundances and
dust grain properties produced by metal-free supernovae with masses 12 - 30 $M_\odot$ and
140 - 260 $M_\odot$ shows that the thermal evolution of protostellar coulds is qualitatively the
same (see Schneider et al. 2006).

\begin{figure*}
\center{{
\epsfig{figure=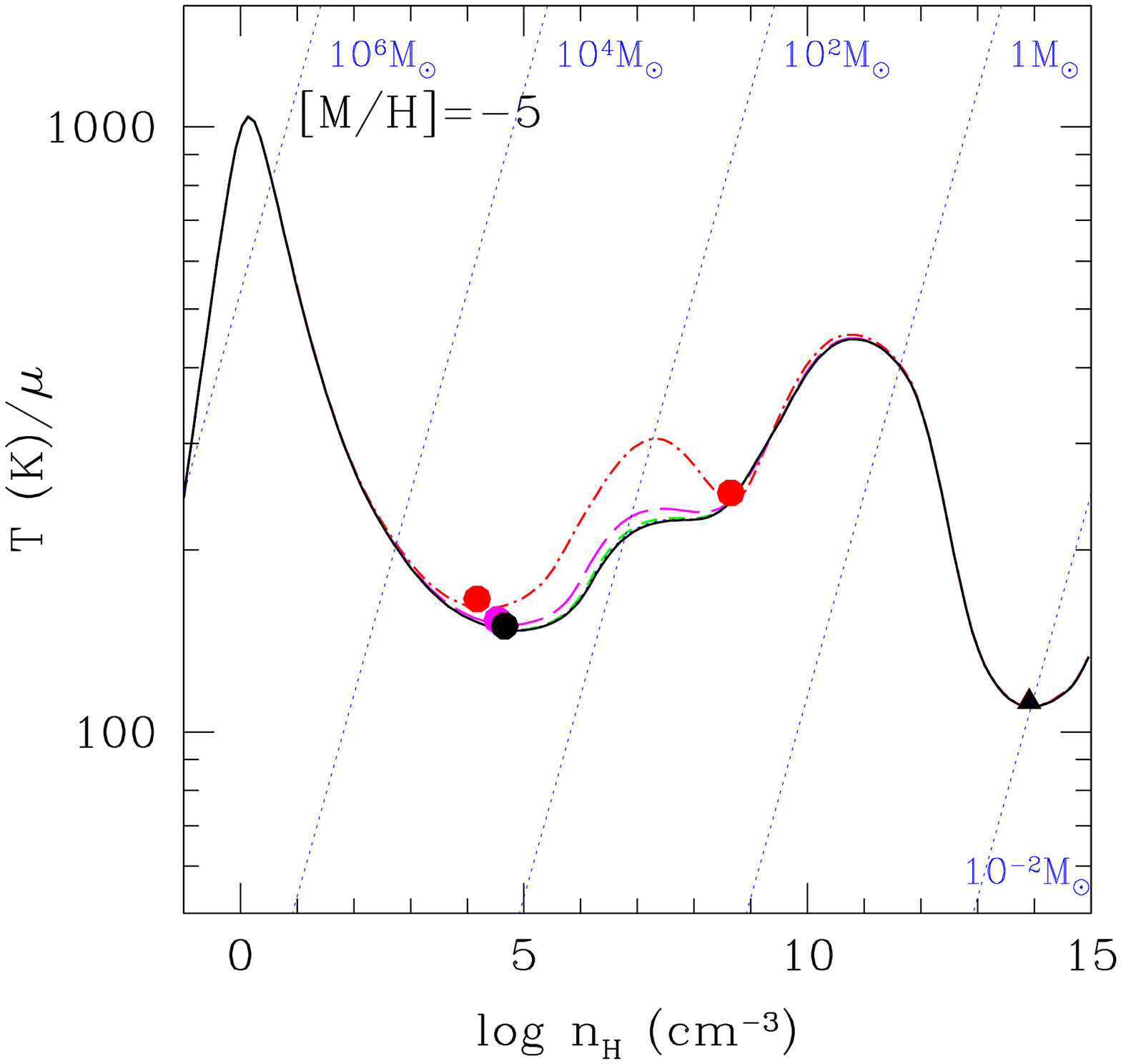,height=6.5cm}
\hspace{1.0cm}
\epsfig{figure=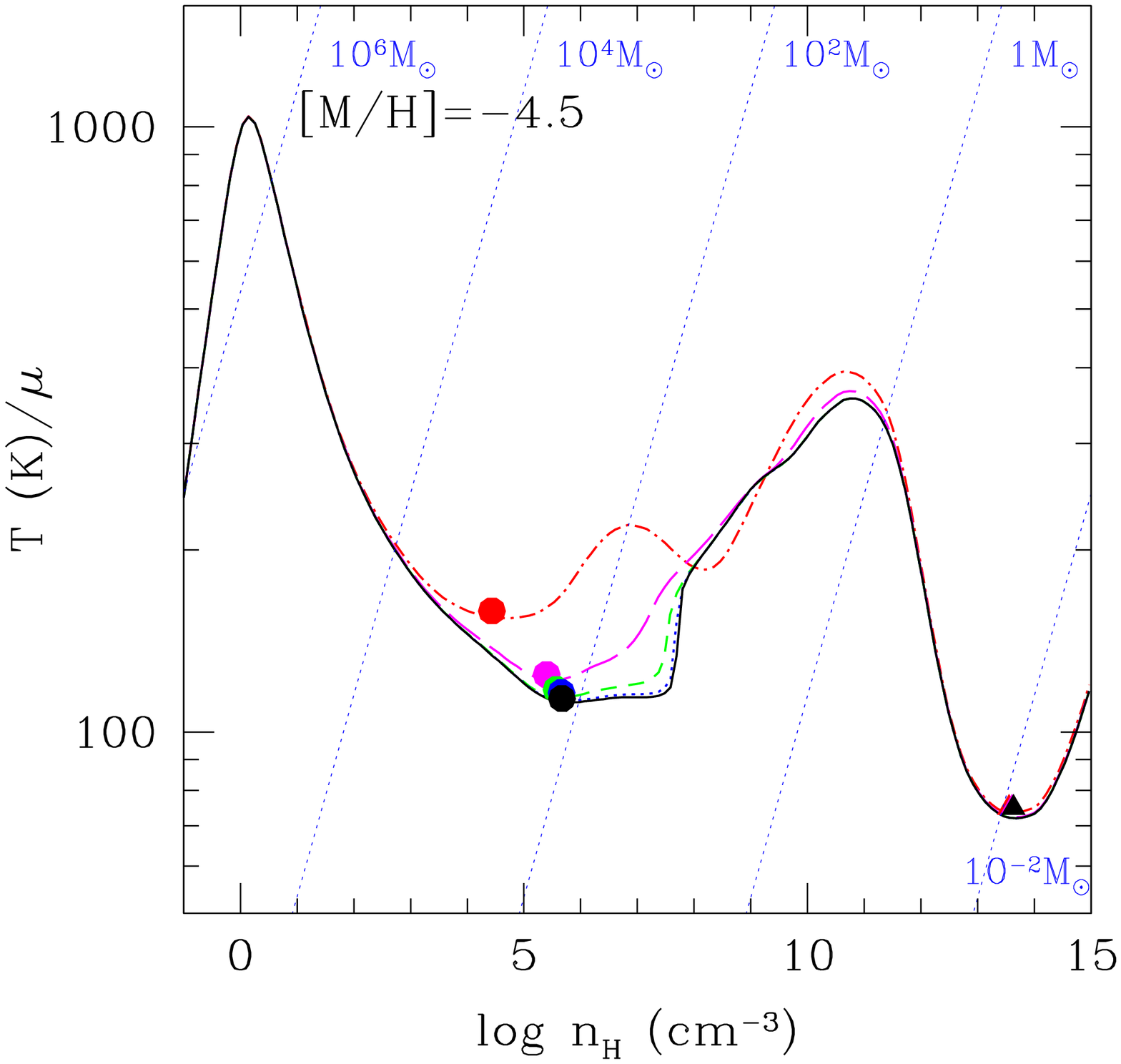,height=6.5cm}
\epsfig{figure=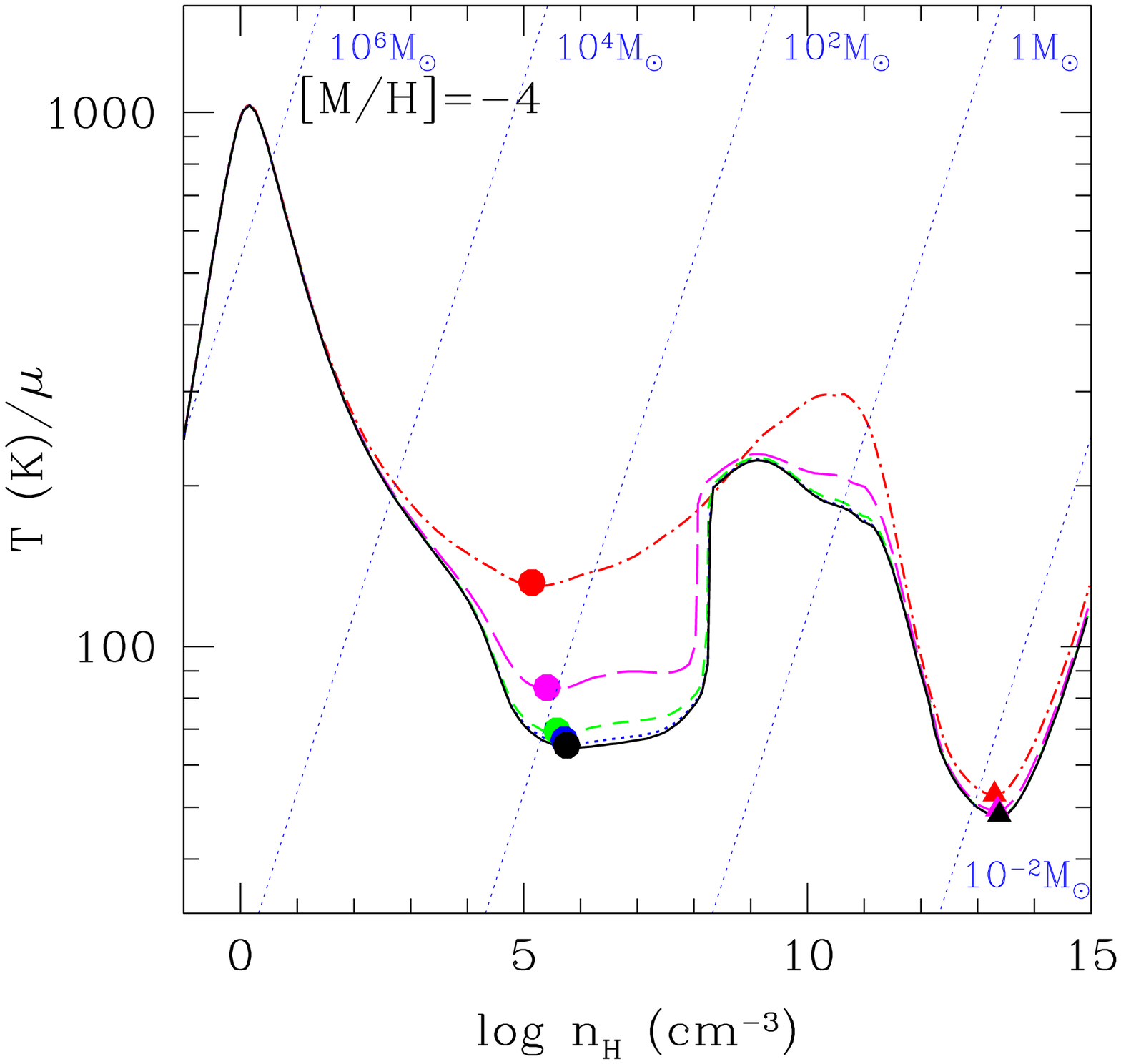,height=6.5cm}
\hspace{1.0cm}
\epsfig{figure=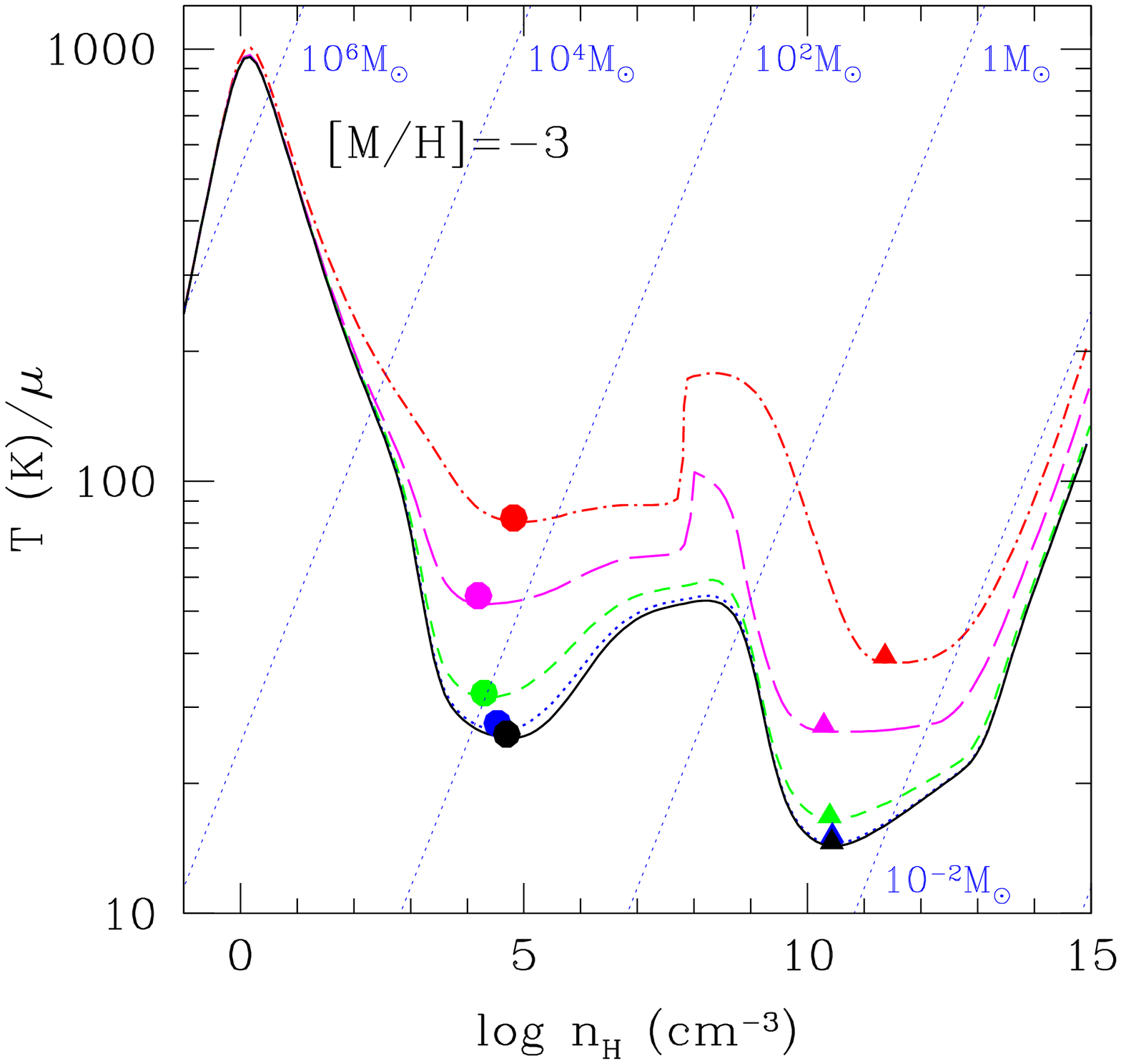,height=6.5cm}
\epsfig{figure=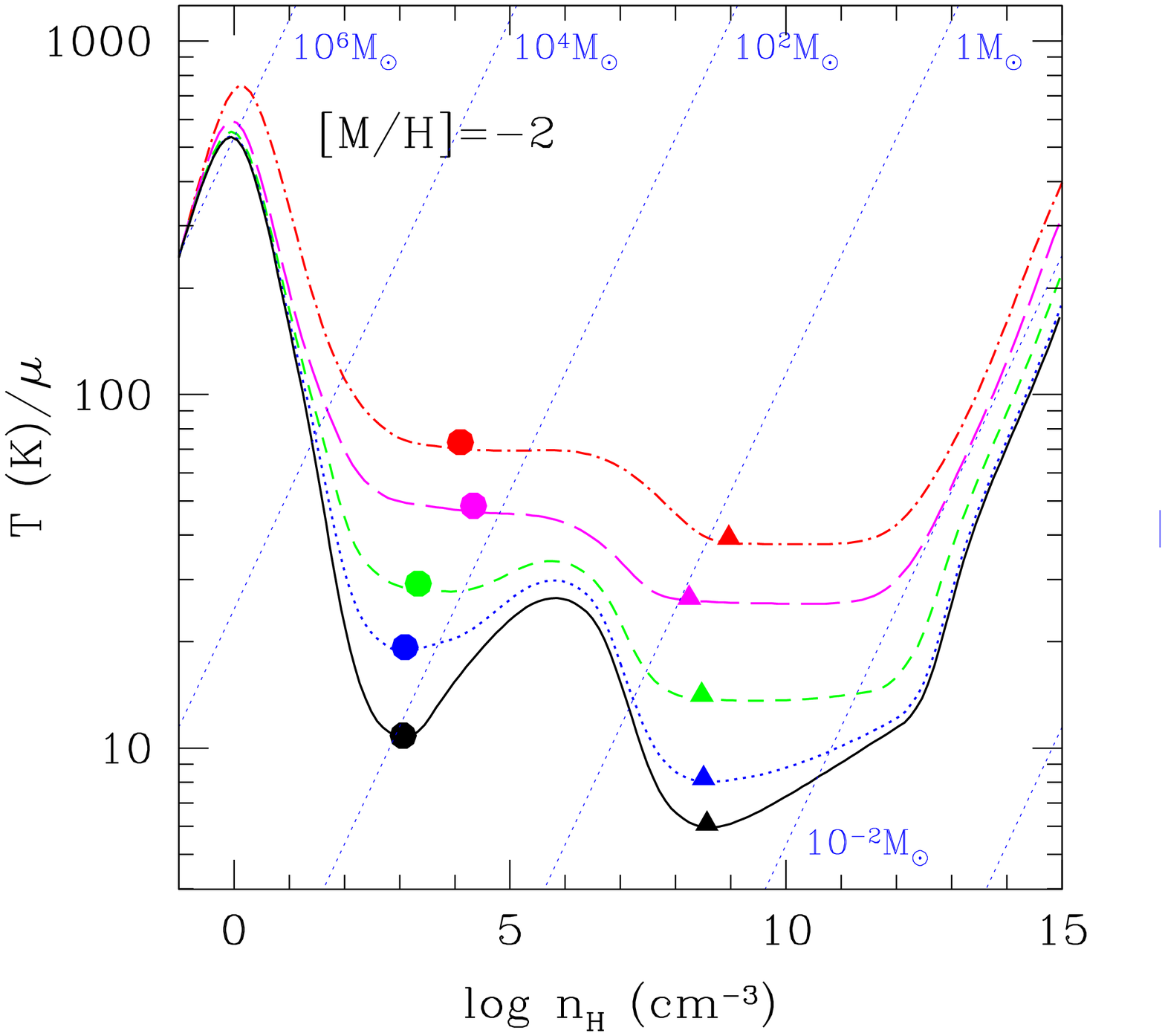,height=6.5cm}
\hspace{1.0cm}
\epsfig{figure=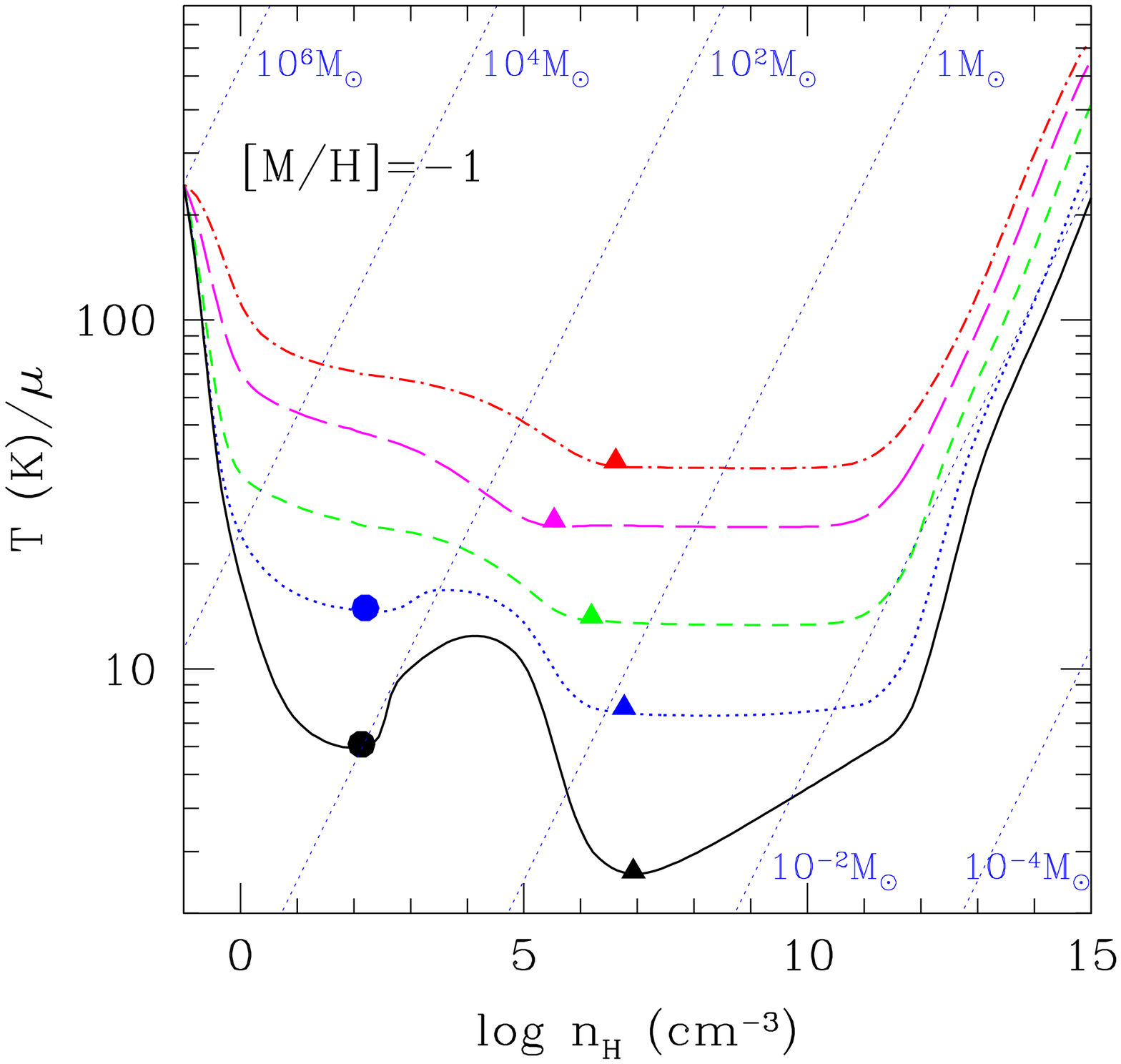,height=6.5cm}
}}
\caption{Thermal evolution as a function of the central gas number density
for protostellar coulds with initial metallicities [M/H]= -5, -4.5, -3, -2, -1 (from the
top-left to the bottom-right). Temperature divided by mean molecular weight is shown.
For our assumed helium abundance ($y_{\rm He}=0.0833$), 
the mean molecular weight changes from $\mu=1.23$ (fully atomic 
hydrogen) to $2.29$ (fully H$_2$). 
In each panel, the different
curves correspond to the evolution of clouds at $z=0$ (solid), 5 (dotted),
10 (short-dashed), 20 (long-dashed), and 30 (dot-dashed). The points 
identify the epochs of fragmentation induced by line-cooling (dots) 
and dust-cooling (triangles).
The diagonal dotted lines identify values of
the Jeans masses corresponding to given thermal states.}
\label{fig:nT}
\end{figure*}

\section{Results}
\label{sec:results}

In this section we discuss the model predictions for the 
thermal evolution of low-metallicity protostellar clouds at different
redshifts, i.e. exposed to a CMB radiation field at different temperatures.
We then discuss the resulting characteristic fragmentation mass scales.

\subsection{Thermal evolution}

In Fig.~\ref{fig:nT} we show the thermal evolution 
of collapsing protostellar clouds as a function of the
central gas number density. The temperature divided by 
the mean molecular weight $T/\mu$ is shown as its variation 
with density is directly related to the effective 
adiabatic index $\gamma -1 =d{\rm ln}(T/\mu)/d{\rm ln} n_{\rm H}$.
Each panel represents the
evolution for a specific initial metallicity, ranging
from [M/H] $= - 5$ (top-left) to [M/H] $= - 1$ (bottom-right).
For each metallicity, we have explored the
variation with redshift and the different curves in
each panel represent clouds collapsing at $z=0$ 
(solid), 5 (dotted), 10 (short-dashed), 20 (long-dashed), and 30 (dot-dashed).
Thus, the effects of the CMB can be quantified by the deviations
of the curves from the reference solid curve because at $z=0$ 
the CMB temperature is too low to have any appreciable consequence
on the clouds thermal evolution.

As expected, the effects of the CMB increase with redshift 
(i.e. higher radiation temperature) and metallicity 
(i.e. lower gas temperature). 
However, even at metallicities as small as [M/H] $= -5$, the thermal evolution
at $z = 30$ shows deviations from the $z = 0$ track, despite the fact that
at this low metallicity the gas temperature is higher than the CMB 
value. The deviation is caused by the suppression of H$_2$ formation
on dust grains due to the increased dust temperature, which is almost equal to
the CMB temperature. The H$_2$ formation rate coefficient on dust grains is
expressed as (Tielens \& Hollenbach 1985),

\[
k_{\rm H_2, gr} = \frac{6 \times 10^{-17} {\rm cm^3 s^{-1}} (T/300 {\rm K})^{1/2} 
(Z/Z_{\odot}) f_{\rm a}}{1+4\times 10^{-2} (T+T_{\rm gr})^{1/2} + 
2 \times 10^{-3} T + 8\times 10^{-6} T^2},
\]   
\noindent
with
\[
f_{\rm a} = \frac{1}{1+{\rm exp}[7.5\times10^2(1/75-1/T_{\rm gr})]}.
\]
\noindent
Thus, when $T_{\rm gr} \ga 75~{\rm K}$, $f_{\rm a}$ becomes very small. If the
H$_2$ formation rate is not reduced at high dust temperatures, as suggested
by Cazaux \& Tielens (2004), the dependence of the thermal evolution on the
CMB at these low metallicities would disappear. 

For initial metallicities [M/H] $< -3$, the effects of the CMB on the cloud 
thermal evolution are limited to densities $n_{\rm H} < 10^8$~cm$^{-3}$, in
the regime where line-cooling dominates. 
At higher metallicities, the minimum temperature induced by dust cooling 
reaches the CMB value.
Thus, the deviations from the $z = 0$ track are more pronounced and 
are present at all densities $n_{\rm H} \ge 10^3$cm$^{-3}$, 
up to the regime where cooling is dominated by dust grains. 
Note that at [M/H] $\ge -2$ and $z \ga 20$,  decrease in $T/\mu$ beyond the 
point where the temperature reaches the CMB floor is due to the increase in 
$\mu$ as a result of H$_2$ formation. 

In each panel, the points identify the epochs of fragmentation induced 
by line-cooling (dots) and by dust-cooling (triangles), while the
diagonal dotted lines represent some indicative values of the Jeans mass
corresponding to the thermal states. As expected on the basis of
arguments given in section 1, the fragmentation epochs correspond to the 
inflection points of the equation of state. 

\begin{figure}
\center{{\epsfig{figure=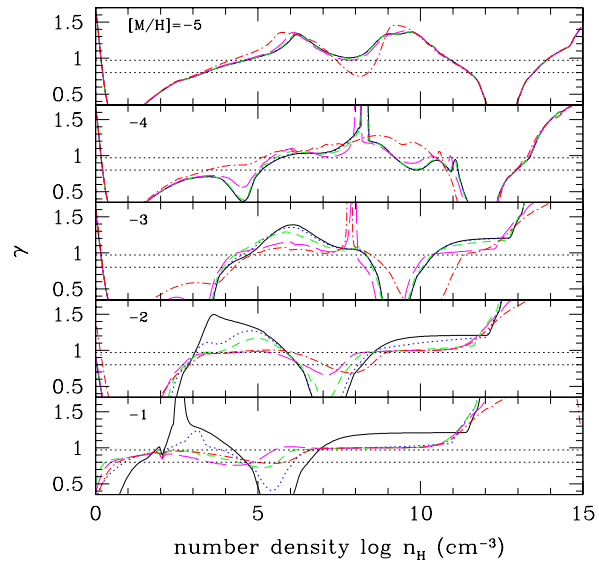, height=8.5cm}}}
\caption{Evolution of the adiabatic index $\gamma$ for clouds with
initial metallicities [M/H] = $ -5, -4, -3, -2, -1$ (from top to bottom). 
In each panel, the different curves show the evolution for a fixed metallicity and redshift $z=0$
(solid), 5 (dotted), 10 (short-dashed), 20 (long-dashed), and 30 (dot-dashed).
The horizontal dotted lines indicate the values limiting the fragmentation
epoch, i.e. a cooling phase with $\gamma < 0.8$ followed by the condition that
$\gamma > 0.97$ (see text).}
\label{fig:gamma}
\end{figure}

More quantitatively, we identify the fragmentation epochs by the condition that
the adiabatic index becomes $\gamma > 0.97$ after a phase of cooling
(where $\gamma << 1$). The choice of a threshold value slightly smaller 
than unity is motivated by the fact that when the gas reaches the CMB 
temperature, $\gamma$ may approach unity asymptotically, but never 
exceeds it. As an additional condition for fragmentation, we impose 
that $\gamma < 0.8$ during the cooling phase preceding the 
fragmentation epoch: this is to eliminate "false" fragmentation points
where $\gamma$ is less than 1 only for a short period of time. 
The evolution of the adiabatic index $\gamma$ as a function of
the central number density for the same set of models shown in
Fig.~\ref{fig:nT} is shown in Fig.~\ref{fig:gamma}.     

For all but the lowest (highest) metallicity models, the two fragmentation phases
(line-induced, at low densities, and dust-induced at high-densities) are preserved 
even in the presence of the CMB. When [M/H] $= -5$ and $z = 30$, however,
there appears to be an intermediate fragmentation phase which is due to H$_2$O cooling
at densities $10^7-10^8$cm$^{-3}$. At lower redshifts/higher metallicities, this 
phase is no longer present because at these intermediate densitites the temperature is
already too low (due to H$_2$/metal line-cooling) for H$_2$O cooling to have any 
appreciable effect. However, as it can be inferred from the behaviour of the dot-dashed
line in the top panel of Fig.~\ref{fig:gamma}, the intermediate fragmentation phase is 
most probably artificial as it depends very much on the fragmentation condition. With
a stricter condition on the duration of the cooling phase, such as $\gamma < 0.7$, it would
not be classified as a fragmentation phase. 
 
Finally, as it can be inferred from the bottom panels of Figs.\ref{fig:nT} and
\ref{fig:gamma}, when [M/H] $> -2$ and $z \ge 10$, the line-induced 
and dust-induced fragmentation phases merge and the thermal evolution is controlled by the 
CMB temperature.  

\subsection{Fragmentation mass}

\begin{figure*}
\center{{
\epsfig{figure=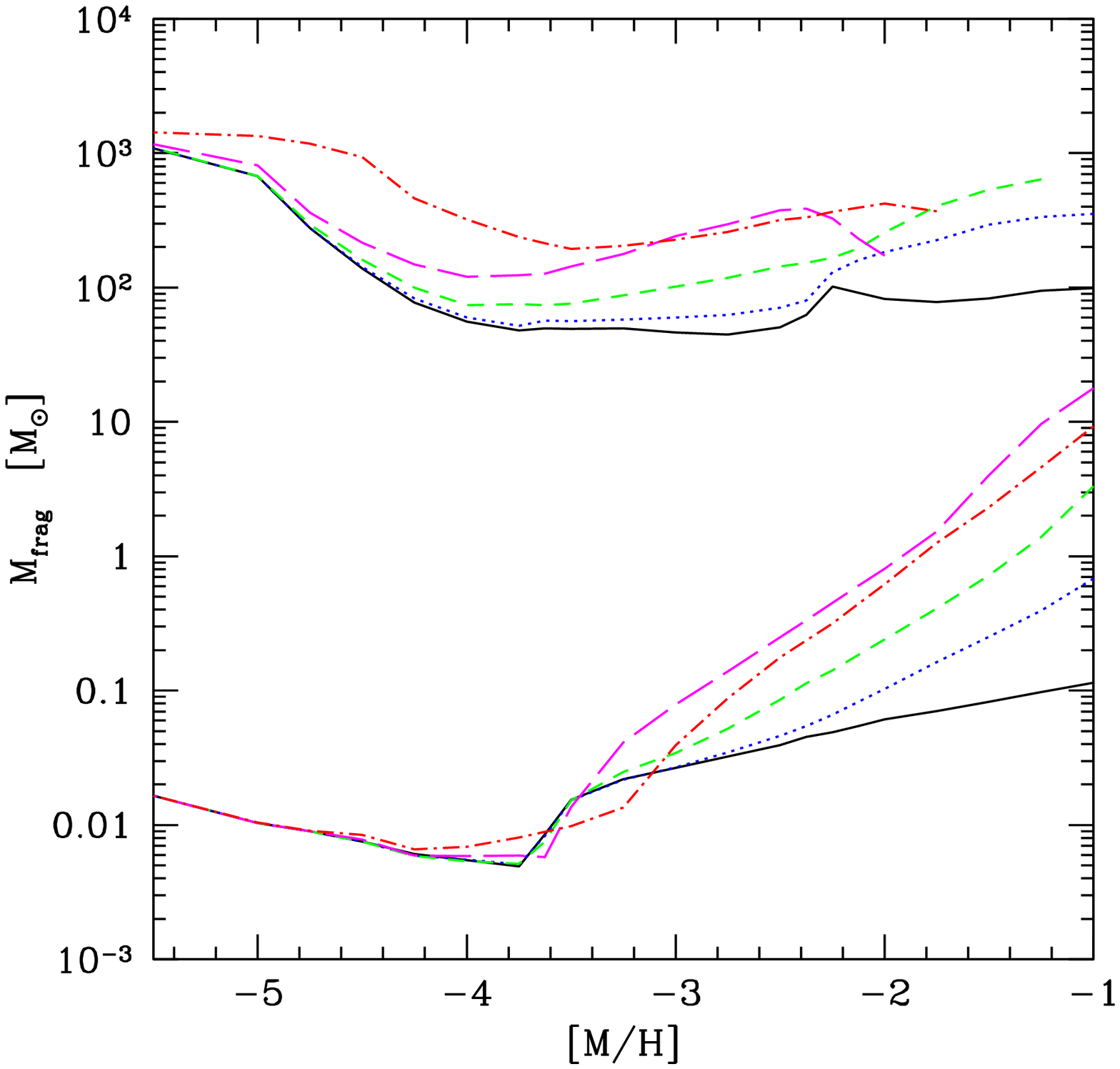,height=8.0cm}
\hspace{1.0cm}
\epsfig{figure=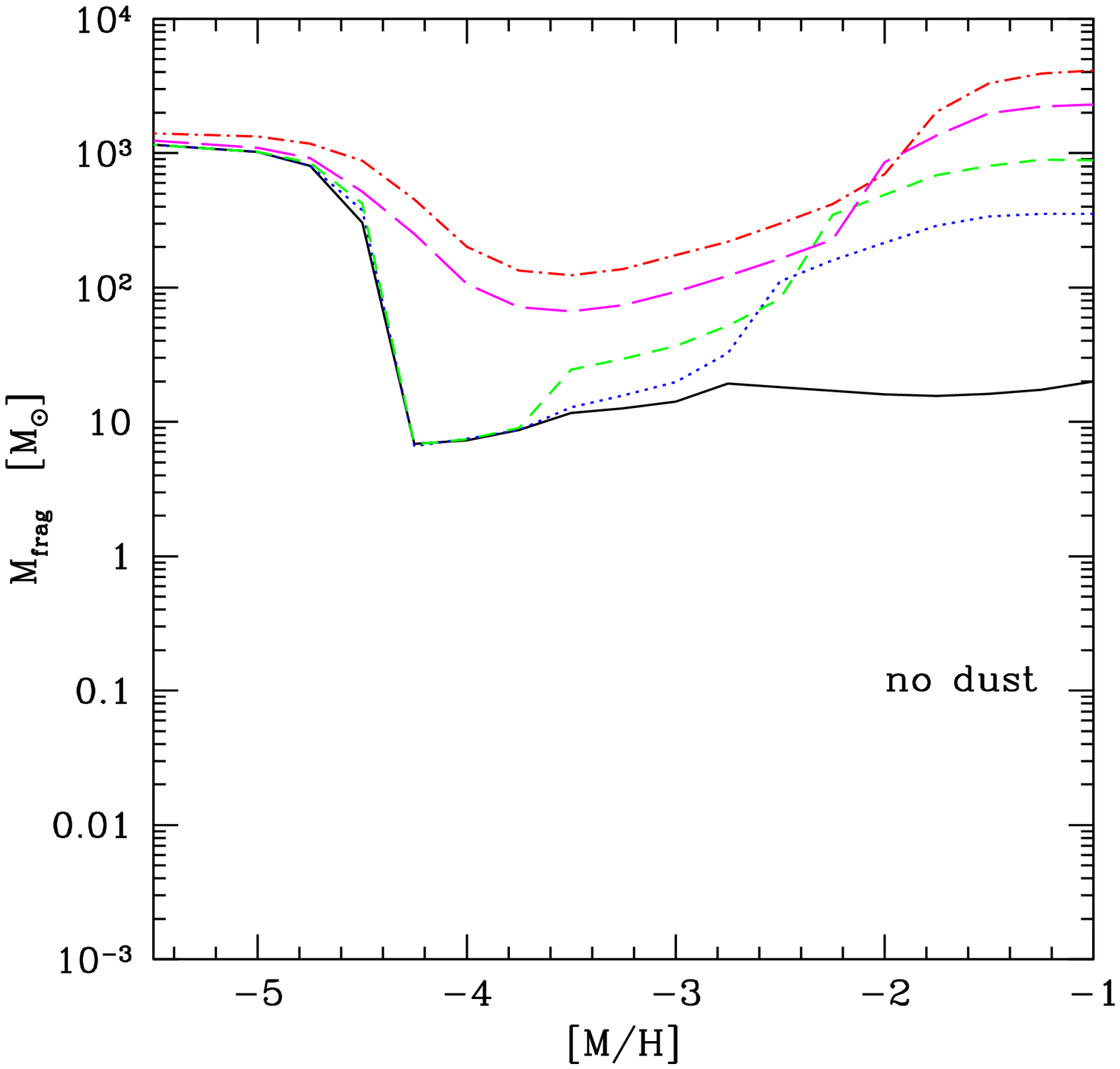,height=8.0cm}
}}
\caption{Characteristic fragmentation mass as a function of metallicity for $z=0$ (solid), 5 (dotted), 
10 (short-dashed), 20 (long-dashed), and 30 (dot-dashed). The left panel shows the same set of 
models presented in Fig.~\ref{fig:nT} with the lower (upper) branch representing the masses associated 
to dust- (metal-) induced fragmentation. The right panel illustrates a test model with no dust.}
\label{fig:mfrag}
\end{figure*}

In the present context, where we assume that thermal pressure is the main force opposing gravity,
the characteristic mass of each fragment is given by the Jeans mass at fragmentation epoch, given in  eq.~\ref{eq:mfrag}.
Fig.~\ref{fig:mfrag} shows the fragmentation masses as a function of metallicity and redshift. In the
left panel, the upper and lower branches are associated to line-induced and dust-induced fragmentation, 
respectively. Since line-induced fragmentation occurs at relatively low densities, the corresponding 
fragment masses are large. In particular, at the lowest metallicities we find that, at any redshift, 
the characteristic fragmentation mass scale, induced by H$_2$ line cooling, is $M_{\rm frag} \sim 10^3 M_\odot$, 
consistent with the results of numerical models. Still, even for metallicities in the range $-5 \le [M/H]  \le -2$, 
metal-line cooling leads to fragment masses $M_{\rm frag} \ge 50 M_\odot$ (Schneider et al. 2006). Dust-induced
fragmentation, which occurs at much higher densities, leads to significantly lower mass scales. 

The CMB affects the amplitude of the characteristic mass in both line- and dust-induced phases.
At any metallicity in the range $-5 \le$ [M/H] $\le -1$, the presence of the CMB leads to larger
characteristic masses in the line-induced phase, with a deviation from the reference $z=0$ track
which increases with metallicity and redshift. Even for clouds with metallicity [M/H] = -4, the 
characteristic mass is more than a factor 2 larger at $z \ge 20$. As it has already been discussed, 
the main effect of the CMB at these low metallicities is the suppression of H$_2$ formation on dust grains. 
In the dust-induced phase, instead, the effects of the CMB on the characteristic mass start to be 
appreciable only when the metallicity is [M/H] $\ge -3$, when the temperature dip due to dust cooling 
hits the CMB floor. At [M/H] $= -3$, the deviation of the characteristic mass from the $z = 0$ track is
by more than a factor 2 at $z \ge 20$. At the highest metallicity considered in this analysis, [M/H] $= -1$,
the characteristic mass in the line- (dust-) induced fragmentation is 100 (0.1), 350 (0.7), 640 (3.3) $M_\odot$
at $z = 0, 5, 10$, respectively. At $z \ge 20$, it is no longer possible to discriminate between the two
fragmentation phases (this is why the long-dashed and dot-dashed lines in the upper branch are interrupted)
and the fragmentation masses are $\approx 10 M_{\odot}$.       

The above results are therefore consistent with the general trend suggested by Smith et al. (2009), that 
the characteristic fragment mass $M_{\rm frag}$ first is high, then it decreases, and then
it is high again (3 modes) but in our analysis this trend is much less pronounced. To better compare
with their analysis, we have run a set of test models with no dust and the characteristic masses 
in the line-induced fragmentation branch as a function of metallicity 
and redshift are shown in the right panel of Fig.~\ref{fig:mfrag}. 

Since the total metallicity is the same in models with and without
dust, about a factor of three more metals are present in the gas-phase 
in the no-dust cases.  
At $z=0$, $M_{\rm frag} \approx 1000 M_{\odot}$
when the metallicity is [M/H] $< -4$ and $M_{\rm frag} 
\approx 10 M_{\odot}$ above this threshold. 
At $z \le 10$, the CMB starts to affect the characteristic mass
when [M/H] $\ge -3.5$, increasing it to several 10s (100s) $M_{\odot}$ when [M/H] $= -3$ ($\ge -2$). At higher
redshifts, the characterstic fragment masses are $\approx 1000 M_{\odot}$ when [M/H] $< -4$ and [M/H] $> -2$
but never decrease below 100 $M_{\odot}$ in the intermediate range $-4 \le $ [M/H] $ \le -2$. Therefore, 
our analysis confirm that the CMB inhibits fragmentation at large metallicities but, in the absence of
dust grains, we do not find any metallicity/redshift range where fragments of a few solar masses can form. As
already pointed out in section 1, in their simulation Smith et al. (2009) do not consider H$_2$ formation
heating nor H$_2$ formation on dust grains which may explain why they tend to find smaller fragment masses 
in the intermediate metallicity range. Note also that their model includes more atomic 
species than just C and O, such as Fe, Si, and S, which provide additional cooling.    

\section{Conclusions}
\label{sec:conclusions}

In this paper we have investigated the dependence of thermal and fragmentation
properties of star-forming clouds on the environmental conditions, focussing on
the interplay between metals, dust grains and the CMB at different redshifts.

Our results, based on an extended grid of models with varying initial metallicity
and redshift, can be summarized as follows:

\begin{enumerate}

\item In the absence of dust grains, when fragmentation occurs at the end of the line-cooling phase,
moderate characteristic masses (of 10s of $M_{\odot}$) are formed at redshifts $z \le 10$ when the metallicity 
is $10^{-4} Z_{\odot} \le Z \le 10^{-2.5} Z_{\odot}$; at lower metallicities, the clouds follow
the primordial fragmentation mode and at higher metallicity, $Z > 10^{-2.5} Z_{\odot}$, the CMB inhibits
fragmentation and only very large masses $\sim$ 100s of $M_{\odot}$, are formed.
These effects become even more dramatic at $z > 10$, when the fragmentation mass scales are always $\ge$ 100s of
$M_{\odot}$, independent of the initial metallicity. 

\item When dust grains are present, at $z=0$ two regimes are present: the line-induced fragmentation mode, which leads
to fragment masses $> 50 M_{\odot}$ at any metallicity, and the dust-induced fragmentation mode, which occurs at higher densities and leads
to sub-solar mass fragments when $Z \ge 10^{-6} Z_{\odot}$. 
The effects of the CMB at $z > 0$ are important in both physical regimes: at very low metallicities, $Z < 10^{-4} Z_{\odot}$, line-cooling
becomes less efficient because of the suppression of H$_2$ formation on dust grains when the grain temperature becomes $\ge 75$~K, at $z > 20$.
At higher metallicities, metal line-cooling causes the gas to thermally couple with the CMB and the resulting fragments are very massive, with $M_{\rm frag} > 100 M_{\odot}$
for $z \ge 5$. Dust-cooling remains relatively insensitive to the presence of the CMB up to metallicities of $Z \sim 10^{-3} Z_{\odot}$ and sub-solar mass 
fragments are formed at any redshift. Above this threshold, heating of dust grains by the CMB at $z \ge 5$ favors the formation of larger masses, which become super-solar 
when $Z \ge 10^{-2} Z_{\odot}$ and $z \ge 10$. 

\item Clouds enriched by metals and dust grains with a total metallicity $Z \ge 10^{-1} Z_{\odot}$ collapsing at $z > 5$,
experience only a single fragmentation epoch (line-induced and dust-induced phases merge) and the characteristic masses
are controlled by the CMB temperature, with $M_{\rm frag} = 0.5, 3, 20 M_{\odot}$ for $z = 10, 20, 30$.   

\end{enumerate}  

\begin{figure*}
\center{{
\epsfig{figure=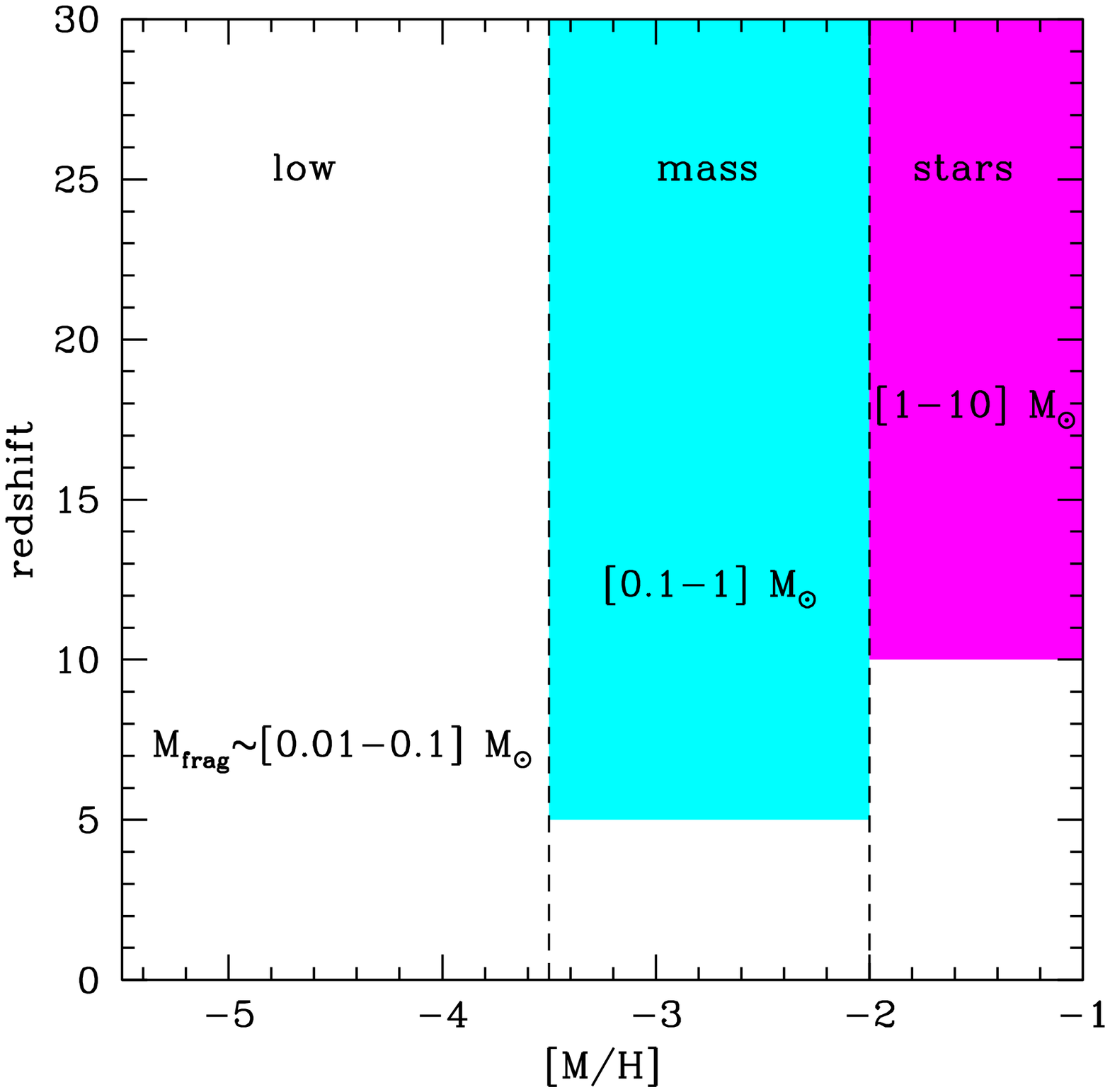,height=8.0cm}
\hspace{1.0cm}
\epsfig{figure=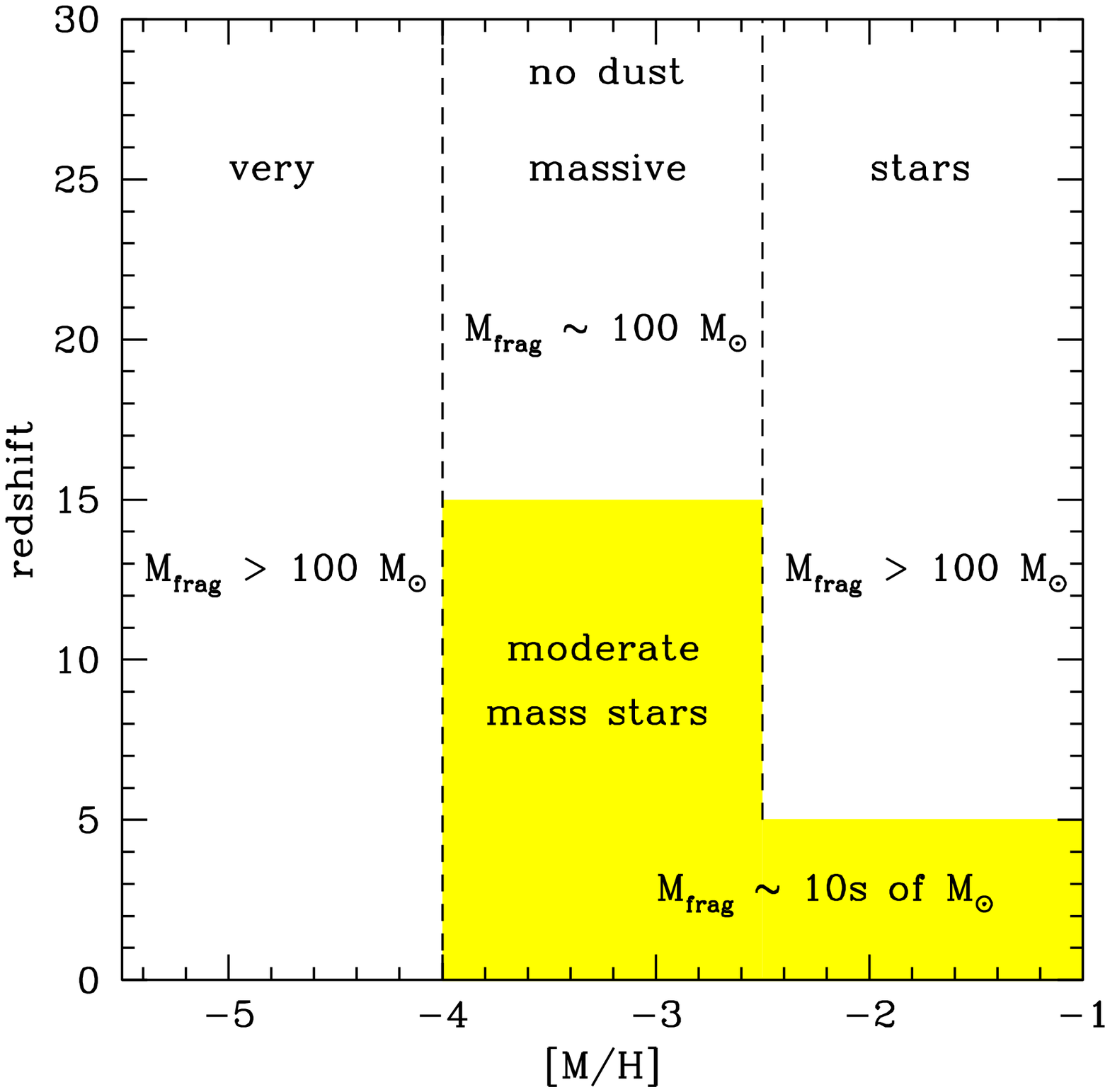,height=8.0cm}
}}
\caption{Schematic represtentation of the characteristic masses associated to the relevant fragmentation regimes
in the metallicity-redshift plane.
In the left panel, we report only the typical scales associated to dust-induced fragmentation, since line-induced
fragmentation leads to masses $> 50 M_{\odot}$ independent of the initial metallicity or redshift. In the right
panel, we show the scales expected in models with no dust (see text).}
\label{fig:summary}
\end{figure*}

The characteristic fragmentation masses are schematically summarized 
in Fig.~\ref{fig:summary}, where we report the typical scales associated to dust-induced fragmentation
(left panel) and in the absence of dust (right panel) in the different redshift-metallicity regimes 
discussed above. We have not considered the fragmentation masses associated to line-cooling in the models
with dust as they are always $> 50 M_{\odot}$ independent of the initial metallicity and redshift (see the upper
branch in the left panel of Fig.~\ref{fig:mfrag}).

Our results confirm the general trend suggested by previous semi-analytic (Omukai et al. 2005)
and numerical models (Smith et al. 2009), that the CMB tends to inhibit fragmentation of relatively 
metal-enriched clouds collapsing at high redshift. However, the interplay between metals, dust and
the CMB generates a complex dependence of the fragmentation mass on environmental conditions.

It is important to stress that in order to explore the observational consequences of such
an environmental dependence, it would be necessary to infer the variation of the functional 
form of the IMF and not only of the characteristic mass. This issue has been recently addressed
by Clark et al. (2009) who predict that, in star forming regions with $Z > 10^{-5} Z_{\odot}$,
competitive accretion can provide a natural explanation for the power-law form of the IMF,
which is observed to be remarkably uniform in a wide variety of environments. The competitive 
accretion of gas from a common reservoir by a collection of protostellar cores requires that the
characteristic fragmentation mass be much smaller than the mass of the cooling gas available. 
As a consequence, Clark et al. (2009) suggest that there may be a link between the CMB and the
slope of the IMF and that, in the absence of dust, 
competitive accretion will not occurr in regions with metallicities
$Z \ge 10^{-3.5} Z_{\odot}$ collapsing at high redshift whereas it will occurr at all redshifts
and metallicities if dust cooling can operate.

\begin{figure}
\center{{\epsfig{figure=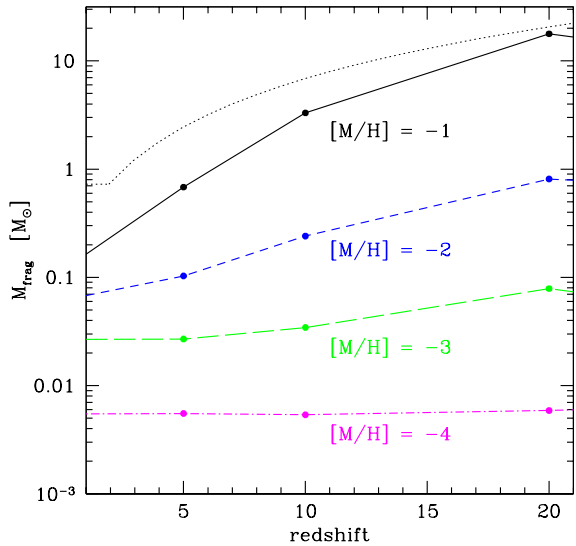, height=8.5cm}}}
\caption{Redshift evolution of the 
fragmentation mass  $M_{\rm frag}$ for clouds with
initial metallicities [M/H] = $-4, -3, -2, -1$ (from bottom to top). 
The thin dotted line shows the empirical dependence of the stellar characteristic
mass on the CMB temperature proposed by Tumlinson (2007, see text).} 
\label{fig:mfrag_red}
\end{figure}

As we have mentioned in section 1, one of the main motivation for the present study was to assess
whether the CMB could be a viable way to modulate the stellar IMF with redshift as suggested
by recent analyses of the elemental abundances of VMP and CEMP stars (Tumlinson 2007; Komiya et al. 2007;
Heger \& Woosley 2008). In Fig.~\ref{fig:mfrag_red} we show the redshift dependence of the fragmentation
mass scales found in the present analysis selecting only the parameter space which is compatible with
fragment masses in the range $M_{\rm frag} \le 10 M_{\odot}$ in the presence of dust grains. 
As a comparison, the thin dotted line shows the 
empirical dependence of the characteristic stellar mass on redshift proposed by Tumlinson (2007, see eq.~1).
It is important to stress that the relation between fragmentation mass scale and characteristic mass shaping the stellar IMF
is not straighforward and several processes, such as competitive accretion or dynamical encounters,
can act to make it not monothonic. However, the inspection of Fig.~\ref{fig:mfrag_red} suggests that for
all but the highest metallicity, the redshift dependence predicted by our analysis is generally shallower 
than what inferred from the empirical IMF. Moreover, observations indicate that the fraction of
CEMP stars decreases with metallicity being 100\% at [Fe/H] $<-4.5$, $\sim 40 \%$ at [Fe/H] $< -3.5$ and
$\sim 20 \%$ at [Fe/H] $< -2$ (Beers \& Christlieb 2005). If CEMP stars reflect the underlying IMF
in the range $1 M_{\odot} < M < 8 M_{\odot}$,  our results seem to indicate an opposite trend: the 
effects of the CMB on the characteristic mass increase with metallicity, suggesting that the CMB can not
be the only physical driver for the required modulation of the stellar characteristic mass.

\section*{Acknowledgments}
We are grateful to the referee, Simon Glover, for his careful reading of the manuscript
which has helped us to improve the quality of the paper. We thank Andrea Ferrara and Stefania
Salvadori for useful comments. RS acknowledges the support and hospitality of the National 
Astronomical Observatory of Japan.

\label{lastpage}
\end{document}